\pdfoutput=1

\documentclass[11pt]{article}
\usepackage[preprint]{acl2016}

\usepackage{acl2016}
\usepackage{times}
\usepackage{latexsym}

\usepackage{multirow}
\usepackage{graphicx}
\usepackage{xcolor,soul}
\usepackage{booktabs}       
\usepackage{xcolor,colortbl}
\usepackage{dashrule}
\usepackage{xspace}
\usepackage{amsmath}
\usepackage{fontawesome}
\usepackage{mdframed}
\usepackage[hang,flushmargin]{footmisc}
\usepackage{enumitem}
\usepackage{amssymb}
\usepackage{natbib}
\usepackage{subcaption}

\usepackage{array}      
\usepackage{graphicx}   
\newcommand\rot[1]{\rotatebox[origin=c]{90}{#1}}   

\newcolumntype{V}{>{\centering\arraybackslash}p{0.3cm}<{\hspace{0pt}}}
\newcolumntype{C}{>{\centering\arraybackslash}p{1.25cm}<{\hspace{0pt}}}

\newcommand{\OmniEmbed}{\texttt{OmniEmbed}\xspace}
\newcommand{\OmniEmbedMultivent}{\texttt{OmniEmbed\textsubscript{Multivent}}\xspace}
\newcommand{\drama}{\texttt{DRAMA-1B}\xspace}
\newcommand{\multvent}{MultiVENT\xspace}

\newcommand{\blue}{\textcolor[HTML]{567AB8}{\bf blue}\xspace}
\newcommand{\orange}{\textcolor[HTML]{F6B26B}{\bf orange}\xspace}

\title{
MAGMaR Shared Task System Description: \\ Video Retrieval with OmniEmbed
}

\author{
Jiaqi Samantha Zhan$^1$\thanks{Equal Contribution\quad \faEnvelopeO~x93ma@uwaterloo.ca}\quad
Crystina Zhang$^{*1}$\quad
Shengyao Zhuang$^{*2,3}$\quad 
\textbf{Xueguang Ma}$^{*1}$\quad \\
\textbf{Jimmy Lin}$^1$\\
[1ex]
 $^1$University of Waterloo \quad $^2$CSIRO \quad $^3$The University of Queensland
}

\date{}

\begin{document}

\maketitle

\begin{abstract}
Effective video retrieval remains challenging due to the complexity of integrating visual, auditory, and textual modalities. In this paper, we explore unified retrieval methods using OmniEmbed, a powerful multimodal embedding model from the Tevatron 2.0 toolkit, in the context of the MAGMaR shared task. Evaluated on the comprehensive MultiVENT 2.0 dataset, OmniEmbed generates unified embeddings for text, images, audio, and video, enabling robust multimodal retrieval. By fine-tuning OmniEmbed with the combined multimodal data—visual frames, audio tracks, and textual descriptions—provided in MultiVENT 2.0, we achieve substantial improvements in complex, multilingual video retrieval tasks. Our submission achieved the highest score on the MAGMaR shared task leaderboard among public submissions as of May 20th, 2025, highlighting the practical effectiveness of our unified multimodal retrieval approach.
Model checkpoint in this work is open-sourced\footnote{\url{https://huggingface.co/Tevatron/OmniEmbed-v0.1-multivent}.}.

\end{abstract}

\section{Introduction}

Information in the real world spans across modalities, including text, images, audio, and videos. While text and image retrieval have advanced considerably due to recent developments in large language models (LLMs) and vision-language models~\cite{bge_dataset, setrank, zhang2025rankwogpt, gme, ma2025tevatron20unifieddocument, zhuang2025documentscreenshotretrieversvulnerable},
video retrieval remains particularly challenging.
This difficulty arises from the complex combination of visual, auditory, and temporal information contained within videos, which traditional text-based methods fail to fully capture.

\begin{figure}[t]
    \centering
    \begin{subfigure}[t]{\linewidth}
        \centering
        \includegraphics[width=\textwidth]{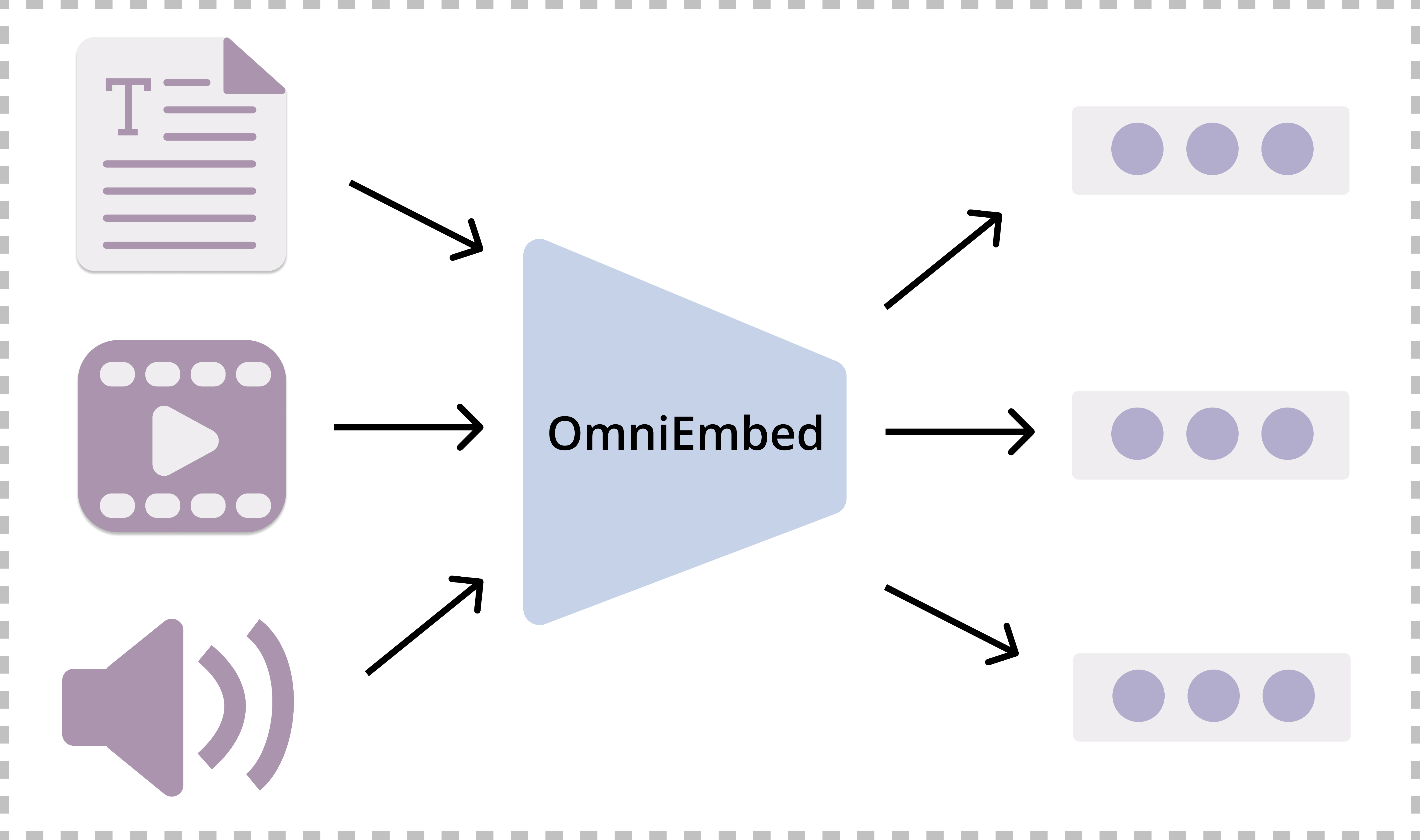}
        \label{fig:fig1}
    \end{subfigure}
    
    \vspace{-0.5em}
    \begin{subfigure}[t]{\linewidth}
        \centering
        \includegraphics[width=\textwidth]{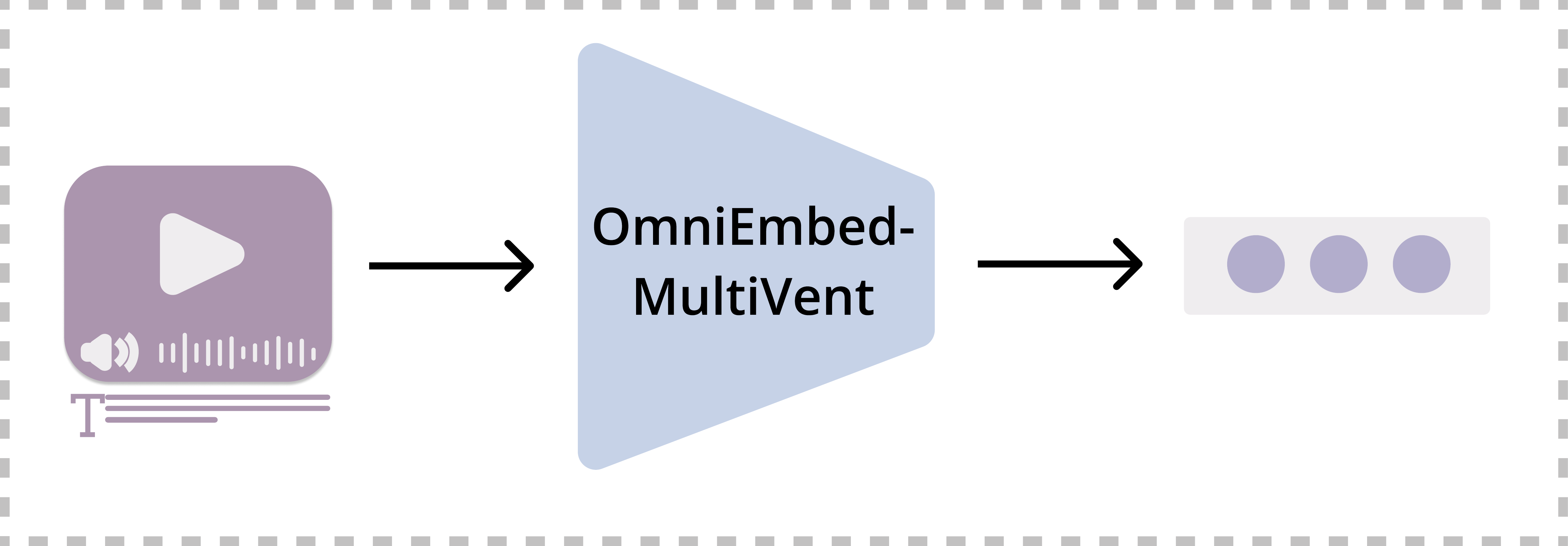} 
        \label{fig:fig2}
    \end{subfigure}
\label{fig:mlutivent}
\vspace{-1em}
\caption{OmniEmbed~\cite{ma2025tevatron20unifieddocument} is originally trained on individual modality content. In this work, we further fine-tune OmniEmbed with MultiVENT 2.0 to enhance its ability to handle compound multimodal information.}
\vspace{-1em}
\end{figure}

Historically, video retrieval has predominantly relied on textual metadata such as captions, descriptions, or extracted textual content via optical character recognition (OCR) and automatic speech recognition (ASR). However, these approaches frequently overlook critical visual and auditory nuances inherent in the video content itself, leading to limited retrieval accuracy and effectiveness.

To address these limitations, MultiVENT 2.0~\cite{kriz2025multivent,sanders2023multivent} was introduced as a comprehensive, multilingual, event-based video retrieval dataset. MultiVENT 2.0 includes over 217,000 videos paired with extensive features such as CLIP~\cite{radford2021clip} and SigLIP~\cite{zhai2023siglip} visual embeddings, Whisper ASR transcripts~\cite{radford2022whisper}, OCR outputs~\cite{paddleocr2020,etter2023multilingualocr}, and video captions generated by Florence~\cite{xiao2023florence}.
By incorporating a diverse range of data, MultiVENT 2.0 provides a realistic and robust benchmark for evaluating advanced video retrieval systems in multilingual contexts.

Concurrently, advancements in large multimodal models (LMMs), exemplified by Omni-LLM architectures like Qwen-2.5-Omni~\citep{Qwen2.5-Omni}, have created opportunities for unified retrieval approaches.
\OmniEmbed, part of the Tevatron 2.0 toolkit~\cite{ma2025tevatron20unifieddocument},
leverages these powerful multimodal models to generate universal embeddings effective across multiple modalities, including text, images, audio, and video. This unified embedding approach simplifies system design, enhances generalization, and facilitates improved multimodal retrieval performance.

In this paper, we use the MultiVENT 2.0 dataset within the MAGMaR shared task to investigate the effectiveness of OmniEmbed-based unified retrieval models. Our goal is to demonstrate how integrating various modalities through end-to-end unified embeddings significantly enhances performance in complex, multilingual video retrieval tasks, reflecting practical real-world applications.

\section{OmniEmbed-Multivent}

\subsection{Initialization}
We initialized our model using OmniEmbed-v0, provided by the Tevatron 2.0 toolkit. OmniEmbed-v0 was trained on diverse multimodal datasets, including text-based retrieval tasks from the BGE dataset~\cite{bge_dataset}, image-document retrieval tasks from ColPali~\cite{faysse2024colpali}, PixMo~\cite{deitke2024molmopixmoopenweights}, and Wiki-SS~\cite{ma2025tevatron20unifieddocument}. Additionally, OmniEmbed-v0 utilized video and audio retrieval training data from MSRVTT~\cite{xu2016msrvtt} and AudioCaps~\cite{dim2019audiocaps}. Despite supporting embeddings across all modalities, the initial OmniEmbed-v0 training did not involve combined multimodal instances. However, real-world videos typically contain multiple combined modalities, including visual content, audio tracks, and descriptive text commonly found on video platforms such as YouTube. To enhance our model's capability for these practical scenarios, we further fine-tuned OmniEmbed using the MultiVENT 2.0 training data, jointly processing textual descriptions, video frames, and audio signals as combined multimodal inputs.

\subsection{Training Data}
\label{sec:experimental_setup:data}
To train the embedding model effectively, it is important to construct meaningful triplets consisting of a query, a positive document, and hard negative documents to optimize the model using the InfoNCE loss~\cite{oord2019representationlearningcontrastivepredictive}.
In traditional text retrieval, hard negatives are typically mined using sparse retrievers like BM25~\cite{bm25}.
However, MultiVENT 2.0 contains a significant number of cross-lingual queries, making BM25 less effective.

To address this, we leverage \drama~\cite{ma2025dramadiverseaugmentationlarge}, a strong multilingual dense retriever. For each training query, we use \drama to encode the associated textual information—specifically, the Whisper-generated ASR transcripts—and retrieve the top 50 documents from the corpus. Any document within this top-50 set that is not labeled as positive is treated as a hard negative. This approach allows us to obtain more semantically relevant and linguistically aligned negatives, which improves the quality of contrastive training in the multilingual setting.

\begin{table*}[t]
\centering
\resizebox{0.9\textwidth}{!}{ 
\begin{tabular}{lllccccc}
\toprule
 & \textbf{Modality} & \textbf{Model} & \multicolumn{1}{l}{\textbf{nDCG@10}} & \multicolumn{1}{l}{\textbf{AP}} & \multicolumn{1}{l}{\textbf{nDCG}} & \multicolumn{1}{l}{\textbf{RR}} & \multicolumn{1}{l}{\textbf{R@10}} \\
 \midrule
 & \multicolumn{7}{c}{\it Official Baselines} \\
 & All & VAST & 0.116 & 0.08 & 0.115 & 0.198 & 0.118 \\
 & OCR & ICDAR OCR $\rightarrow$ CLIP & 0.217 & 0.166 & 0.288 & 0.363 & 0.227 \\
 & ASR & Whisper ASR & 0.267 & 0.212 & 0.336 & 0.417 & 0.29 \\
 & Vision (key frame) & CLIP & 0.304 & 0.261 & 0.435 & 0.429 & 0.333 \\
 & All & LanguageBind & 0.324 & 0.283 & 0.452 & 0.443 & 0.355 \\
 \midrule
 & \multicolumn{7}{c}{\it Zero-Shot} \\
(a) & text, ASR & \drama & 0.629 & 0.576 & 0.693 & 0.749 & 0.649 \\
(b) & text, ASR & \OmniEmbed & 0.377 & 0.329 & 0.453 & 0.493 & 0.403 \\
(c) & text, ASR, Vision (video), Audio & \OmniEmbed & 0.595 & 0.537 & 0.673 & 0.732 & 0.616 \\
 \midrule
 & \multicolumn{7}{c}{\it Trained on MultiVent 2.0 Training Set} \\
(d) & text, ASR & \OmniEmbedMultivent & 0.710 & 0.673 & 0.772 & 0.808 & 0.734 \\
(f) & Vision (video), Audio & \OmniEmbedMultivent & 0.709 & 0.665 & 0.776 & 0.822 & 0.724 \\
(h) & text, ASR, Vision (video), Audio & \OmniEmbedMultivent & 0.753 & 0.769 & 0.807 & 0.848 & 0.715 \\
\bottomrule
\end{tabular}
}
\caption{
    Overall scores on the test set of MultiVENT 2.0.
    The indices are not continuous to match \autoref{tab:multivent_test_breakdown}.
}
\vspace{-0.5cm}
\label{tab:multivent_test_overall}
\end{table*}

\section{Experimental Setup}

\subsection{Training}
We perform continual LoRA fine-tuning on \texttt{Tevatron/OmniEmbed-v0.1}, which uses the Thinker module of Qwen2.5-Omni-7B as its backbone. Each training query is paired with one positive document and three hard negative documents.
For each video input, we uniformly sample 24 frames, which are fed into the model together with audio and textual features. Both training and inference is conducted via the Tevatron toolkit~\cite{tevatron}.

\subsection{Evaluation}
Evaluation is conducted on the test queries of the MAGMaR shared task using MultiVENT 2.0, which contains 2,549 test queries and 109,800 test documents in the corpus. Model effectiveness is measured by submitting the retrieval results to Eval.ai, where nDCG@10 serves as the primary evaluation metric.
\section{Results}

\subsection{Overall Results}
\autoref{tab:multivent_test_overall} compares the average scores of our methods along with the official baselines provided by \citet{kriz2025multivent}.
We start with the text-only baselines:\
\drama, which exhibits strong multilingual retrieval capability (row~a), 
and zero-shot \OmniEmbed, which is trained on multi-modal data but only infers on text descriptions (row~b). 

\paragraph{Zero-shot text-only results.}
We find that the two text-only baselines already yield visible improvement over the provided baselines, which are based on single- or multi-modality channels.
We also notice the importance of incorporating diverse multilingual training data.
While row~(b) employs a 7B model, it yields much inferior results compared to the 1B \texttt{DRAMA} model on row~(a),
which has been heavily exposed to multilingual data. 

\paragraph{Zero-shot multi-modality results.}
With a strong base of text-only retrieval capacity, we move on and explore whether vision and audio information provide additional gains. 
Surprisingly, our multimodal retriever exhibits much greater generalization ability by adding raw video and audio data without any post-processing (e.g., keyframe extraction, OCR, or ASR):
Taking nDCG@10 as an example, the overall score is improved by 21.8 points\footnote{compare row~b to row~c, 0.377 $\rightarrow$ 0.595}.
While it does not surpass \drama on the overall scores, the effectiveness gap is closer,
where the gap mainly lies on the Text-based queries, while the multimodal \OmniEmbed outperforms \drama on the Speech- and OCR-based queries (\autoref{tab:multivent_test_breakdown}, Query-Type section).

Another interesting observation under this experimental setting ---
as there is no other large-scale training data that aligns all three modalities, the model behind row~(b, c) is trained on pairwise modality data.
In other words, it only sees text--text, text--video, or text--audio pairs in each training data sample,
but no text--video--audio triplet.\footnote{As mentioned in Section~\ref{sec:experimental_setup:data}, text--video training data comes from MSR-VTT~\cite{xu2016msrvtt}, and text--audio data comes from AudioCaps~\cite{dim2019audiocaps}.} 
As a result, inference on three modalities together on the test set of \multvent not only enforces ``zero-shot'' on the video length and data domain, but also on the inference methods,
as it requires the model to transfer the relevance pattern separately learned on each modality into a uniform pattern across all modalities.

\paragraph{In-domain results.}
With in-domain training data, both text-only and multi-modal retrievers are greatly improved, with nDCG@10 achieving over 0.7.
Other than the encouraging overall effectiveness boost, we share two insightful observations:\

\textbf{1. Non-text modalities can achieve almost \textit{identical} effectiveness as text-only modality.} 
As mentioned, with high-quality in-domain data, the text-only retriever is greatly improved with nDCG@10 increased by 33.3 points\footnote{compare row~(c) to (d), 0.377 $\rightarrow$ 0.710}.
While the text-only retriever exhibits strong effectiveness,  our retriever based only on vision and audio input is able to catch up and yield almost identical scores on all evaluation metrics\footnote{compare row~(d) to (f)}
\textit{without any text information}.

\begin{table*}[]
\centering

\resizebox{\textwidth}{!}{ 
\begin{tabular}{Vr|CCC|CCCCC}
\toprule

& &
\multicolumn{3}{c}{\bf Zero-shot} & 
\multicolumn{5}{c}{\bf Trained on MultiVent 2.0}
\\
& \textbf{Configuration} & {(a)} & \multicolumn{1}{c}{(b)} & \multicolumn{1}{c}{(c)} & \multicolumn{1}{c}{(d)} & \multicolumn{1}{c}{(e)} & \multicolumn{1}{c}{(f)} & (g) & (h) \\
\cmidrule{1-10}
& \textbf{Backbone} & \drama & \multicolumn{2}{c}{\OmniEmbed} & \multicolumn{5}{c}{\OmniEmbedMultivent} \\
\cmidrule{1-10}

\multicolumn{10}{c}{\emph{Training data}} \\
\cmidrule{1-10}

 & \textbf{text$^*$} & -- & -- & -- & \checkmark & \checkmark & -- & \checkmark & \checkmark \\
 & \textbf{audio} & -- & -- & -- & -- & \checkmark & \checkmark & \checkmark & \checkmark \\
 & \textbf{video} & -- & -- & -- & -- & \checkmark & \checkmark & \checkmark & \checkmark \\
 
\cmidrule{1-10}

\multicolumn{10}{c}{\emph{Inference data}} \\
\cmidrule{1-10}
 & \textbf{text$^*$} & \checkmark & \checkmark & \checkmark & \checkmark & \checkmark & -- & -- & \checkmark \\
 & \textbf{audio} & -- & -- & \checkmark & -- & -- & \checkmark & \checkmark & \checkmark \\
 & \textbf{video} & -- & -- & \checkmark & -- & -- & \checkmark & \checkmark & \checkmark \\
 
\midrule
\multicolumn{10}{c}{\emph{Breakdown Results}} \\
\cmidrule{1-10}

\multirow{6}{*}{\rot{\bf Language}} &
\textbf{Arabic} & \cellcolor[HTML]{FEFFFF}0.440 & \cellcolor[HTML]{FBE0C5}0.243 & \cellcolor[HTML]{ECF4FB}0.504 & \cellcolor[HTML]{C1DAF0}0.684 & \cellcolor[HTML]{C3DBF1}0.676 & \cellcolor[HTML]{C1DAF1}0.682 & \cellcolor[HTML]{C0D9F0}0.688 & \cellcolor[HTML]{B3D1ED}0.742 \\
 & \textbf{English} & \cellcolor[HTML]{EFF6FC}0.500 & \cellcolor[HTML]{FDF3E8}0.351 & \cellcolor[HTML]{FEFDFC}0.416 & \cellcolor[HTML]{FDFEFF}0.435 & \cellcolor[HTML]{F5F9FD}0.465 & \cellcolor[HTML]{FCFDFF}0.437 & \cellcolor[HTML]{FEFFFF}0.428 & \cellcolor[HTML]{F7FBFE}0.457 \\
 & \textbf{Korean} & \cellcolor[HTML]{FCE8D4}0.296 & \cellcolor[HTML]{F9CB9C}0.112 & \cellcolor[HTML]{FCE5CD}0.269 & \cellcolor[HTML]{FDF0E4}0.339 & \cellcolor[HTML]{FDF1E6}0.345 & \cellcolor[HTML]{FDF0E3}0.335 & \cellcolor[HTML]{FDF1E5}0.342 & \cellcolor[HTML]{FDF3E9}0.355 \\
 & \textbf{Russian} & \cellcolor[HTML]{EFF6FC}0.502 & \cellcolor[HTML]{FCECDB}0.312 & \cellcolor[HTML]{FEFEFD}0.419 & \cellcolor[HTML]{E0ECF8}0.555 & \cellcolor[HTML]{DFECF8}0.560 & \cellcolor[HTML]{E9F2FA}0.516 & \cellcolor[HTML]{EBF3FB}0.507 & \cellcolor[HTML]{E0ECF8}0.555 \\
 & \textbf{Chinese} & \cellcolor[HTML]{FBDEC0}0.231 & \cellcolor[HTML]{F9CB9C}0.113 & \cellcolor[HTML]{FCE5CE}0.272 & \cellcolor[HTML]{FCE8D4}0.289 & \cellcolor[HTML]{FCEAD7}0.300 & \cellcolor[HTML]{FCEBDA}0.307 & \cellcolor[HTML]{FCEBD9}0.306 & \cellcolor[HTML]{FDEEE0}0.326 \\
 & \textbf{Spanish} & \cellcolor[HTML]{BCD7EF}0.709 & \cellcolor[HTML]{FAFCFE}0.444 & \cellcolor[HTML]{B3D2ED}0.740 & \cellcolor[HTML]{ABCDEB}0.773 & \cellcolor[HTML]{A6CAEA}0.794 & \cellcolor[HTML]{A7CAEA}0.793 & \cellcolor[HTML]{A6CAEA}0.795 & \cellcolor[HTML]{9FC5E8}0.823 \\
\cmidrule{1-10}

\multirow{6}{*}{\rot{\bf Query Type}} &

\textbf{Base} & \cellcolor[HTML]{FDF4EA}0.612 & \cellcolor[HTML]{FAD8B5}0.390 & \cellcolor[HTML]{FDF5EC}0.620 & \cellcolor[HTML]{F9FBFE}0.706 & \cellcolor[HTML]{ECF4FB}0.723 & \cellcolor[HTML]{EDF4FB}0.722 & \cellcolor[HTML]{F2F7FC}0.715 & \cellcolor[HTML]{D9E8F6}0.749 \\
 & \textbf{Text} & \cellcolor[HTML]{D0E3F4}0.762 & \cellcolor[HTML]{FCEBD9}0.539 & \cellcolor[HTML]{FEFFFF}0.699 & \cellcolor[HTML]{A6C9EA}0.819 & \cellcolor[HTML]{9FC5E8}0.828 & \cellcolor[HTML]{D0E3F4}0.762 & \cellcolor[HTML]{D8E7F6}0.751 & \cellcolor[HTML]{A2C7E9}0.824 \\
 & \textbf{Speech} & \cellcolor[HTML]{FDF6EE}0.629 & \cellcolor[HTML]{FAD3AD}0.355 & \cellcolor[HTML]{FEFBF8}0.670 & \cellcolor[HTML]{E4EFF9}0.733 & \cellcolor[HTML]{D0E3F4}0.761 & \cellcolor[HTML]{BBD6EF}0.790 & \cellcolor[HTML]{BBD6EF}0.791 & \cellcolor[HTML]{B2D1ED}0.802 \\
 & \textbf{OCR} & \cellcolor[HTML]{FCEAD7}0.534 & \cellcolor[HTML]{F9CB9C}0.284 & \cellcolor[HTML]{FDF5ED}0.625 & \cellcolor[HTML]{FEF9F3}0.649 & \cellcolor[HTML]{FEFBF7}0.666 & \cellcolor[HTML]{CEE2F4}0.764 & \cellcolor[HTML]{C9DEF2}0.771 & \cellcolor[HTML]{D3E5F5}0.757 \\
 & \textbf{MultiVENT-Base} & \cellcolor[HTML]{FEFBF8}0.671 & \cellcolor[HTML]{FBDDBE}0.427 & \cellcolor[HTML]{FDEFE1}0.575 & \cellcolor[HTML]{D4E5F5}0.756 & \cellcolor[HTML]{CCE1F3}0.767 & \cellcolor[HTML]{F4F8FD}0.713 & \cellcolor[HTML]{FBFDFE}0.702 & \cellcolor[HTML]{C1DAF1}0.781 \\
 & \textbf{MultiVENT-Specific} & \cellcolor[HTML]{FDEFE1}0.575 & \cellcolor[HTML]{F9D1A9}0.339 & \cellcolor[HTML]{FBE4CB}0.484 & \cellcolor[HTML]{FEF9F4}0.651 & \cellcolor[HTML]{FEFBF8}0.669 & \cellcolor[HTML]{FDF6EE}0.626 & \cellcolor[HTML]{FDF5EC}0.619 & \cellcolor[HTML]{FEFEFE}0.694 \\
\cmidrule{1-10}

\multirow{4}{*}{\rot{\bf Video Type}} &
\textbf{Professional} & \cellcolor[HTML]{D9E8F6}0.443 & \cellcolor[HTML]{FBDCBD}0.212 & \cellcolor[HTML]{C3DBF1}0.472 & \cellcolor[HTML]{BED8F0}0.479 & \cellcolor[HTML]{B6D3EE}0.489 & \cellcolor[HTML]{BCD7EF}0.481 & \cellcolor[HTML]{B1D0ED}0.495 & \cellcolor[HTML]{B6D3EE}0.489 \\
 & \textbf{Edited} & \cellcolor[HTML]{DBEAF7}0.440 & \cellcolor[HTML]{FCE7D2}0.269 & \cellcolor[HTML]{FBFDFF}0.397 & \cellcolor[HTML]{BED8F0}0.478 & \cellcolor[HTML]{B3D1ED}0.493 & \cellcolor[HTML]{A2C7E9}0.515 & \cellcolor[HTML]{A8CBEB}0.507 & \cellcolor[HTML]{9FC5E8}0.519 \\
 & \textbf{Diet-Raw} & \cellcolor[HTML]{FDEFE0}0.309 & \cellcolor[HTML]{FADBBB}0.206 & \cellcolor[HTML]{FBE2C8}0.241 & \cellcolor[HTML]{FEFEFE}0.389 & \cellcolor[HTML]{FFFFFF}0.392 & \cellcolor[HTML]{F8FBFE}0.401 & \cellcolor[HTML]{FEFDFB}0.382 & \cellcolor[HTML]{DAE9F7}0.441 \\
 & \textbf{Raw} & \cellcolor[HTML]{FCE8D4}0.275 & \cellcolor[HTML]{FCE6D0}0.264 & \cellcolor[HTML]{F9CB9C}0.118 & \cellcolor[HTML]{FEFEFE}0.389 & \cellcolor[HTML]{FFFFFF}0.392 & \cellcolor[HTML]{FADBBB}0.205 & \cellcolor[HTML]{F9D3AB}0.162 & \cellcolor[HTML]{FEFEFE}0.391 \\
\cmidrule{1-10}

\multirow{8}{*}{\rot{\bf Event Type}} &
\textbf{Disaster} & \cellcolor[HTML]{FEF6EF}0.615 & \cellcolor[HTML]{F9D3AB}0.366 & \cellcolor[HTML]{FCEBD9}0.535 & \cellcolor[HTML]{E6F0F9}0.716 & \cellcolor[HTML]{DFECF8}0.729 & \cellcolor[HTML]{FCFEFF}0.678 & \cellcolor[HTML]{FEFEFE}0.672 & \cellcolor[HTML]{D6E6F6}0.745 \\
 & \textbf{Political} & \cellcolor[HTML]{FDF2E6}0.583 & \cellcolor[HTML]{F9CB9C}0.307 & \cellcolor[HTML]{FDEFE1}0.565 & \cellcolor[HTML]{FEFCFB}0.658 & \cellcolor[HTML]{FFFFFF}0.673 & \cellcolor[HTML]{FEFCFA}0.656 & \cellcolor[HTML]{FEFCFB}0.658 & \cellcolor[HTML]{F5F9FD}0.691 \\
 & \textbf{Election} & \cellcolor[HTML]{FCE7D3}0.511 & \cellcolor[HTML]{F9CB9C}0.307 & \cellcolor[HTML]{FBDDBF}0.437 & \cellcolor[HTML]{FCE8D3}0.513 & \cellcolor[HTML]{FDEFE1}0.562 & \cellcolor[HTML]{FDF5EC}0.603 & \cellcolor[HTML]{FDF2E7}0.584 & \cellcolor[HTML]{FEF7F0}0.618 \\
 & \textbf{Protest} & \cellcolor[HTML]{FDF2E6}0.583 & \cellcolor[HTML]{FAD4AE}0.377 & \cellcolor[HTML]{FBDFC3}0.454 & \cellcolor[HTML]{EEF5FB}0.702 & \cellcolor[HTML]{ECF3FB}0.707 & \cellcolor[HTML]{FEFBF8}0.647 & \cellcolor[HTML]{FEF9F4}0.633 & \cellcolor[HTML]{DFECF8}0.729 \\
 & \textbf{Sports} & \cellcolor[HTML]{FEFDFC}0.663 & \cellcolor[HTML]{FCE5CD}0.491 & \cellcolor[HTML]{EEF5FB}0.702 & \cellcolor[HTML]{DBEAF7}0.736 & \cellcolor[HTML]{CFE2F4}0.757 & \cellcolor[HTML]{BBD6EF}0.792 & \cellcolor[HTML]{C1DAF1}0.781 & \cellcolor[HTML]{B0CFEC}0.812 \\
 & \textbf{Social} & \cellcolor[HTML]{FEFCFA}0.655 & \cellcolor[HTML]{FAD4AE}0.374 & \cellcolor[HTML]{FEF7F1}0.621 & \cellcolor[HTML]{DEEBF7}0.732 & \cellcolor[HTML]{D3E5F5}0.750 & \cellcolor[HTML]{E9F2FA}0.712 & \cellcolor[HTML]{E8F1FA}0.714 & \cellcolor[HTML]{B6D3EE}0.800 \\
 & \textbf{Science} & \cellcolor[HTML]{FEFEFF}0.676 & \cellcolor[HTML]{FAD5B0}0.381 & \cellcolor[HTML]{FFFFFF}0.673 & \cellcolor[HTML]{D8E8F6}0.741 & \cellcolor[HTML]{C2DAF1}0.780 & \cellcolor[HTML]{BED8F0}0.787 & \cellcolor[HTML]{C2DAF1}0.780 & \cellcolor[HTML]{B8D4EE}0.798 \\
 & \textbf{OtherEvent} & \cellcolor[HTML]{E1EDF8}0.725 & \cellcolor[HTML]{FBE2C9}0.474 & \cellcolor[HTML]{D6E7F6}0.744 & \cellcolor[HTML]{B8D4EE}0.798 & \cellcolor[HTML]{B1D0ED}0.809 & \cellcolor[HTML]{AACCEB}0.821 & \cellcolor[HTML]{ADCEEC}0.817 & \cellcolor[HTML]{9FC5E8}0.841 \\
\midrule

\multicolumn{10}{c}{\emph{Overall Results}} \\
\cmidrule{1-10}
& \textbf{nDCG@10} & \cellcolor[HTML]{FDF2E7}0.629 & \cellcolor[HTML]{F9CB9C}0.377 & \cellcolor[HTML]{FCEDDD}0.595 & \cellcolor[HTML]{F9FCFE}0.710 & \cellcolor[HTML]{D5E6F5}0.727 & \cellcolor[HTML]{FBFDFE}0.709 & \cellcolor[HTML]{FEFEFE}0.705 & \cellcolor[HTML]{9FC5E8}0.753 \\
 & \textbf{AP} & \cellcolor[HTML]{FDF1E5}0.576 & \cellcolor[HTML]{F9CB9C}0.329 & \cellcolor[HTML]{FCEBDA}0.537 & \cellcolor[HTML]{EBF3FB}0.673 & \cellcolor[HTML]{CADFF3}0.691 & \cellcolor[HTML]{F8FBFE}0.665 & \cellcolor[HTML]{FEFEFD}0.657 & \cellcolor[HTML]{9FC5E8}0.715 \\
 & \textbf{nDCG} & \cellcolor[HTML]{FDF2E6}0.693 & \cellcolor[HTML]{F9CB9C}0.453 & \cellcolor[HTML]{FDEEE0}0.673 & \cellcolor[HTML]{FBFDFF}0.772 & \cellcolor[HTML]{D1E3F4}0.788 & \cellcolor[HTML]{F1F7FC}0.776 & \cellcolor[HTML]{FEFEFE}0.769 & \cellcolor[HTML]{9FC5E8}0.807 \\
 & \textbf{RR} & \cellcolor[HTML]{FDF4EA}0.749 & \cellcolor[HTML]{F9CB9C}0.493 & \cellcolor[HTML]{FDF1E5}0.732 & \cellcolor[HTML]{FEFDFC}0.808 & \cellcolor[HTML]{E2EDF8}0.825 & \cellcolor[HTML]{ECF3FB}0.822 & \cellcolor[HTML]{EAF3FA}0.822 & \cellcolor[HTML]{9FC5E8}0.848 \\
 & \textbf{R@10} & \cellcolor[HTML]{FDF3E8}0.649 & \cellcolor[HTML]{F9CB9C}0.403 & \cellcolor[HTML]{FDEDDE}0.616 & \cellcolor[HTML]{E4EFF9}0.734 & \cellcolor[HTML]{C8DEF2}0.749 & \cellcolor[HTML]{FAFCFE}0.724 & \cellcolor[HTML]{FEFEFE}0.718 & \cellcolor[HTML]{9FC5E8}0.769 \\

\bottomrule
\end{tabular}
}
\caption{
    Breakdown scores on the test set of MultiVENT 2.0.
    $^*$:\ Text is a concatenation of video title, video caption, video description, and whisper caption. 
    Under ``Breakdown Results'', entries are highlighted at the Type level. For example, all entries under ``Language'' are highlighted in the same color scale, and all entries under ``Query Type'' are highlighted in another color scale, and so on.
    Under ``Overall Results'', entries are highlighted at the row level.
    In all cases, \blue indicates higher values and \orange indicates lower values.
}
\label{tab:multivent_test_breakdown}
\end{table*}

\textbf{2. Non-text modalities can provide \textit{additional augmentation} of text-only modality.}
While non-text modalities yield strong results, they serve more than a replacement for text information:\
By combining all three modalities together, the retriever yields another 4 points of improvement, from around 0.71 to 0.753, as shown by row~(h).

\smallskip
To the best of our knowledge, it is the first time that non-text modalities alone have achieved similar effectiveness to their text-only counterparts in an end-to-end system.
This surely results from the strong multi-modality capability of our base model, 
and shows that the current multi-modal infrastructure in the community allows us to explore more possibilities of retrieval beyond text.

\subsection{Breakdown Results}
\autoref{tab:multivent_test_overall} reports the average scores over all metrics, and all in-domain results have the same training and inference configurations. 
Through \autoref{tab:multivent_test_breakdown}, we provide more experimental ablations and look at the results through the lens of different types of queries.

\paragraph{Inference modality determines the effectiveness pattern.}
Under in-domain results, we explore the model behavior when they are trained and inferred under different configurations. 
On column~(e) and (g),
while the model is trained on all modalities,
They are only exposed to text-only or non-text modalities at the inference time, respectively.
This leads to an interesting observation that experiments with similar inference configurations show more similar effectiveness patterns under each type category, regardless of the training configuration.
That is, columns~(d) and (e) show similar patterns,\footnote{
    For example, strong effectiveness on the ``Text'' category under ``Query Type'', ''Disaster'' category under ``Video Type'', etc.
}
and columns (f) and (g) show similar patterns.\footnote{
    For example, weaker effectiveness on ``Raw'' category under ``Video Type'', strong effectiveness on ``Sports'' category under ``Event Type'', etc.
}
This suggests our multimodal retriever is relatively robust to the modality mismatch between training and inference time.

\paragraph{When does adding non-text modality help?}
The overall results above suggest that adding multimodal inputs is beneficial in both zero-shot and in-domain settings, but on which scenarios are they beneficial? 
This section shares the patterns observed regarding query and video types,\footnote{
It would be ideal to analyze the query-level categorization, but unfortunately, it is not yet available to the public at the time of writing.
We hope to perform more in-depth analysis in this direction when the resources are available.} 

\textbf{Query Type.}
We observe an aligned pattern on inference modality vs query type,
that text-only retrievers (columns d, e) favor the text-based queries, non-text retrievers (column f, g) favor the speech- and OCR-based queries,
and the full-modality retriever (column h) performs the best on the base queries.

\textbf{Video Type.}
Among all four video types, models perform better on the ''Professional'' and ``Edited'' videos, likely because these videos provide comprehensive information in all text, video, and audio channels.
Interestingly, models achieve visibly better results on ``Diet-Raw'' videos when provided full modalities.
This greatly match the nature of the Diet-Raw videos, which are ``with minimal text and speech overlays''~\cite{kriz2025multivent}.
On the other hand,
``Raw'' categories heavily depend on the text features.
We suspect it is because the video and audio are very noisy under this category and only provide sparse information.

\section{Related Works}

Previous work in video retrieval has largely focused on extracting textual information from other modalities, such as using OCR~\cite{paddleocr2020,etter2023multilingualocr} for embedded text and ASR systems such as Whisper~\cite{radford2022whisper} for audio transcripts.
Alongside text-based approaches, researchers have explored extracting visual and audio features directly from videos using pretrained vision or audio models.~\cite{cao-etal-2024-rap,Croitoru_2021_ICCV}
%
To the best of our knowledge, we are the first to demonstrate an effective end-to-end video retrieval model that fully leverages combined visual, audio, and textual information and is able to outperform traditional text-only algorithms on challenging, multilingual, and event-centric video retrieval benchmarks.

\section{Conclusion}
In this work, we present a unified multimodal retrieval framework using OmniEmbed, fine-tuned on the comprehensive MultiVENT 2.0 dataset for event-centric, multilingual video retrieval. Our results demonstrate that leveraging combined visual, audio, and textual modalities in an end-to-end system substantially outperforms traditional text-only baselines, both in zero-shot and in-domain scenarios. Notably, we show that non-text modalities alone can achieve comparable effectiveness to text, and their combination further boosts retrieval performance. These findings highlight the potential of unified multimodal models for robust, real-world video retrieval applications.

\section*{Acknowledgments}
We thank the MAGMaR organizers for coordinating the shared task and workshop. This research was supported in part by the Natural Sciences and Engineering Research Council (NSERC) of Canada.

\bibliography{acl2016,customize}

\end{document}